# Watching a GaN Transistor Switch: Real-Time Nanoscale Strain and Heat Dynamics

Cedric Corley-Wiciak[1*], Nicolas-Til Sammler[2], Boris Butej[2], Manuel Petersmann[2], Agnieszka A. Corley-Wiciak[1], Laura Neumann[3], Juan M. Carrillo-Larrea[1], Paul Schmiedeke[4], Marc Strätgen[5], Paul-Antoine Douissard[1], Steven J. Leake[1], Peter Boesecke[1], Markus Sievers[2], Peter Imrich[2], Michael Nelhiebel[2], Michael Glavanovics[2], Dionyz Pogany[6], Michael Reisinger[2*], Tobias U. Schulli[1]

[1]ESRF – The European Synchrotron; 38000 Grenoble, France.

[2]KAI GmbH - Kompetenzzentrum Automobil- und Industrieelektronik; 9524 Villach, Austria.

[3]IKTS - Fraunhofer-Institut für Keramische Technologien und Systeme; 01277 Dresden, Germany.

[4]Infineon Technologies AG; 85579 Neubiberg, Germany.

[5]Infineon Technologies Austria AG; 9500 Villach, Austria.

[6]Institute of Solid State Electronics, TU Wien; 1040 Vienna, Austria.

*Corresponding authors. Email: corley@esrf.fr, Michael.Reisinger@k-ai.at

Digital and energy technologies depend on microelectromechanical and power electronic components whose performance is critically impacted by rapid, cyclic deformations. Real-time information on their operation has remained inaccessible due to the need for nanosecond and nanometer resolution in fully integrated devices. We break this limitation by imaging the complete switching cycle of an industrial GaN high electron mobility transistor through stroboscopic dark field X ray microscopy at a fourth-generation synchrotron, resolving electromechanical and thermal micro strain fields across the entire device and correlating them with time dependent voltage characteristics. Coupled simulations benchmarked against the measurements reproduce electric field evolution and transient thermal hotspots. This combined approach provides direct insight into device physics and informs design strategies for next generation energy and information processing technologies.

The transition of our economy to a more sustainable and interconnected future relies heavily on advanced functional materials, such as wide-bandgap semiconductors [1, 2], which enable efficient power conversion in electric vehicles [3], industrial systems [4, 5], and emerging microelectronic platforms [6, 7]. During many such applications, devices are subjected to rapidly varying electric fields and high power densities [8], generating electromechanical deformation and transient thermal hotspots that critically influence performance and reliability [9, 10]. Today, the sub-microscopic dynamics governing these processes remain largely inaccessible in operating devices generally, as conventional characterization tools cannot probe buried active regions with the required nanometer spatial and nanosecond temporal resolution at high frequency [11, 12]. As a result, the development of next-generation materials-based power electronics is largely reliant on theoretical models whose predictions cannot be directly validated [13].

The recent advent of fourth-generation synchrotron sources with unprecedented brightness has enabled a transition from static strain imaging by X-ray diffraction microscopy to recording real-time movies of lattice dynamics at kHz- MHz repetition rates [14, 15, 16]. Here we leverage stroboscopic Dark-Field X-ray Microscopy (DFXM) [17], applied to an industrial (Al)GaN high-electron-mobility transistor (HEMT) as a case study to directly visualize electromechanical and thermal strain evolution in a functional device during its OFF/ON switching cycle [18, 19]. An image sequence with ~100 nm and ~1 ns resolution reveals the spatio-temporal distribution of compressive and tensile deformation inside the active region, while simultaneous electrical measurements allow correlation of strain transients with the build-up of electric fields and power dissipation. Comparison with coupled Technology Computer-Aided Design (TCAD) and Finite Element Method (FEM) simulations yields an experimentally validated model of the electrical, thermal, and mechanical processes that govern operation under realistic working conditions. These results establish *operando* X-ray microscopy as a powerful approach for uncovering nanoscale mechanisms that broadly limit the performance and reliability of materials and devices, extending well beyond microelectronics to batteries, catalysts, and other systems driven by transient external stimuli.

# HEMT Pre-Characterization

**Fig. 1a** shows a top-view optical micrograph of the HEMT, consisting of four transistor structures connected in parallel, with a source-drain distance of 10 µm and a gate width of ~350 µm. The area investigated in this work, indicated by the shading, covers nearly the entire active area of the device.

**Figure 1: HEMT device layout and (Al)GaN/Si heterostructure**

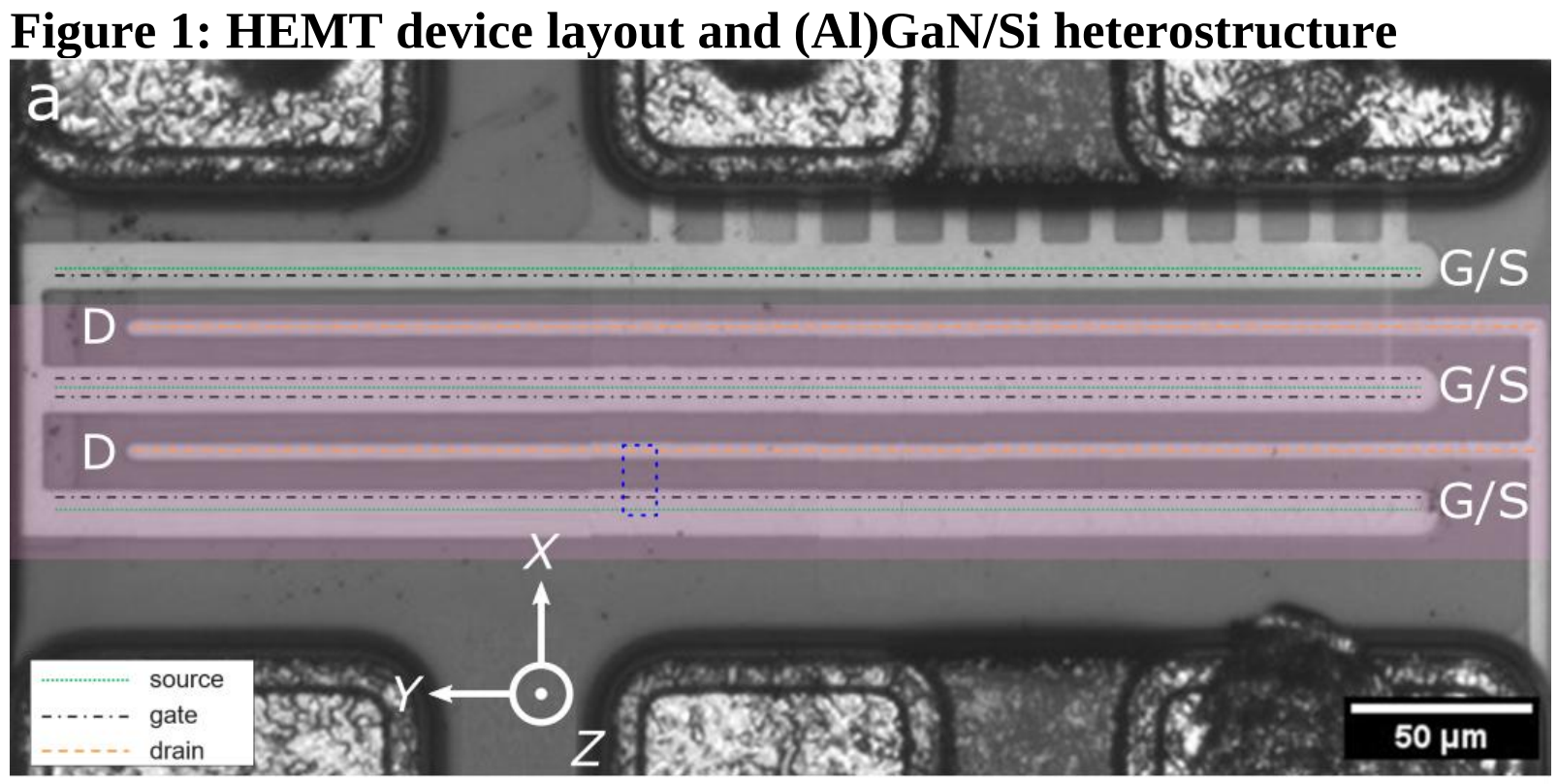

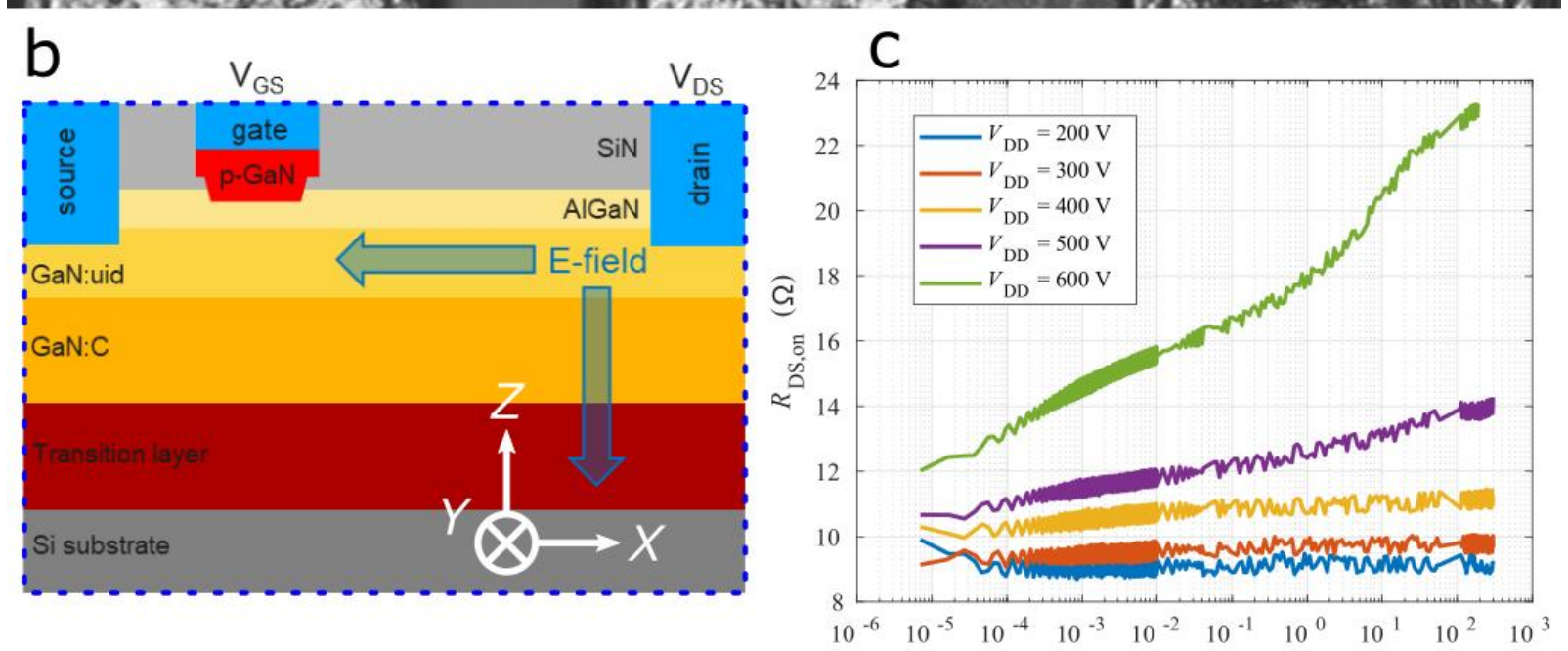

**a,** Optical micrography in top-view of the HEMT, the extents of gate, source, and drain electrodes are indicated by the dashed lines, the pink shading indicates the area measured by DFXM, the dashed blue box indicates the cross-section shown in **b,** Sketch of the epitaxial heterostructure with arrows depicting the lateral and vertical electric fields during OFF-state. **c,** ON-state resistance vs. hard-switching operation time for different drain supply voltages $V_{DD}$.

The epitaxial (Al)GaN heterostructure, sketched in **Fig. 1b**, is grown on a Si(111) substrate and consists of a ~2 µm thick transition layer accommodating lattice and thermal mismatch between Si and GaN [20], a ~1 µm thick carbon-doped GaN layer (GaN:C) layer with a doping concentration of roughly $10^{19}$ cm$^{-3}$, a ~400 nm unintentionally doped GaN (GaN:uid) channel, and a ~60 nm thick $Al_{0.2}Ga_{0.8}N$ barrier. The metallic gate, source, and drain electrodes are fabricated lithographically with a TiN contact layer. The sample is protected from environmental influences by a SiN passivation layer.

The GaN HEMT is subjected to switching with a capacitive load (see Supporting Figure 1). The drain supply voltage was set to $V_{DD}$ = 400 V, while the gate-source voltage was switched between $V_{GS}$ = 0 V (OFF) and $V_{GS}$ = 4 V (ON) at a frequency $f$ = 355 kHz and a duty cycle $DC$ = 20%. As shown in **Fig. 1c**, $V_{DD}$ = 400 V was selected such that the global ON-state resistance $R_{DS,on}$ remains approximately constant throughout the experiment, while it increases over time for higher $V_{DD}$. Thus, the chosen operating point minimized changes associated with trapping, allowing to focus on intrinsic physical processes during switching.

# Stroboscopic Strain Imaging

**Fig. 2a** depicts the time-dependent waveforms of drain-source voltage $V_{DS}(t)$, $V_{GS}(t)$, and drain current $I_D(t)$ during a single switching cycle. In the OFF-state ($V_{GS}$ = 0 V) the device blocks with $V_{DS} \approx V_{DD}$ = 400 V. During turn-ON starting at $t_{OFF}$ = -580 ns, $V_{GS}$ is switched to 4 V and $I_D$ rises, reaching a maximum of ~200 mA at $t \approx$ -490 ns, while simultaneously $V_{DS}$ gradually reduces, reaching the ON-state value $V_{DS} \approx 0$ at $t_{ON}$ = -390 ns. Subsequently, $I_D$ decreases to a plateau of

~40 mA during the ON-state. At $t$ = 0, $V_{GS}$ is switched back to 0 V and the device turns OFF, with $V_{DS}$ gradually returning toward 400 V and $I_D$ decreasing to zero. The strain dynamics induced in the semiconductor material by switching the voltage are captured through stroboscopic DFXM at the hard X-ray microscopy beamline ID01/ESRF [21], with the experimental setup sketched in **Fig. 2b**. Using compound refractive lenses (CRLs) as imaging objective, measurement of the 0002 Bragg reflection probes the out-of-plane lattice strain $\Delta\varepsilon_{zz}$ at each position ($X$, $Y$) on the sample, integrated along the combined thickness of the GaN:uid and GaN:C sublayers. The switching of $V_{GS}(t)$ is synchronized to the repetition frequency of the X-ray pulses, which have a length of ~ 40 ps, and the detector gating, enabling to image the device at a series of points during the OFF/ON cycle by scanning the delay between electrical switching and X-ray probe. In total, a 3D dataset $\varepsilon_{zz}(t, x, y)$ is obtained (see Supporting Figure 3), from which we define the differential strain $\Delta\varepsilon_{zz}(t, x, y)$ as relative change with respect to $t$ = 0:

$$\Delta\varepsilon_{zz}(t, x, y) = \varepsilon_{zz}(t, x, y) - \varepsilon_{zz}(t = 0, x, y) \qquad (1)$$

**Fig. 2c** presents a selection of twelve representative images extracted from the full switching movie of the transistor (see Supporting Figure 4 for all 64 frames). These ‘snapshots’ capture key moments in the spatio-temporal strain evolution, with compressed/tensile regions depicted in blue and red, respectively.

There are two pronounced dynamic phenomena present at each of the four parallel regions: (1) A compressive strain field of $\Delta\varepsilon_{zz} \approx -2 * 10^{-4}$ in the area between gate and drain, which is steady during the OFF-state, and (2) a tensile strain of up to $\Delta\varepsilon_{zz} \approx +1.5 * 10^{-4}$ by the gate edge during turn-ON. By correlating the two types of dynamic strain fields with the time-dependent electronic waveform recorded simultaneously, we can attribute (1) to the IPE response of the GaN lattice from the change of the electrical field ($\Delta\varepsilon_{zz,IPE}$ ~ $V_{DS}$(t)) [12], and (2) to localized Joule heating ($\Delta\varepsilon_{zz,T}$ ~ $I_D$(t)) [9]:

$$\Delta\varepsilon_{zz}(t, x, y) = \Delta\varepsilon_{zz,IPE}(t, x, y) + \Delta\varepsilon_{zz,T}(t, x, y) \qquad (2)$$

In the following, we describe the exact dynamics and origins of both effects.

## Electro-Thermal Dynamics

During OFF-state, we expect a two-dimensional (2D) electrical field $\vec{E} = [E_x \quad 0 \quad E_z]$ in the area between gate and drain. Since the experiment probes the differential out-of-plane strain $\Delta\varepsilon_{zz}$, which is unaffected by the lateral field component $E_x$ [32], we focus the discussion on the vertical field $E_z$. In a wurtzite lattice structure, this component is linked to the corresponding strain

**Figure 2: Spatio-temporal strain imaging by DFXM**
**I**

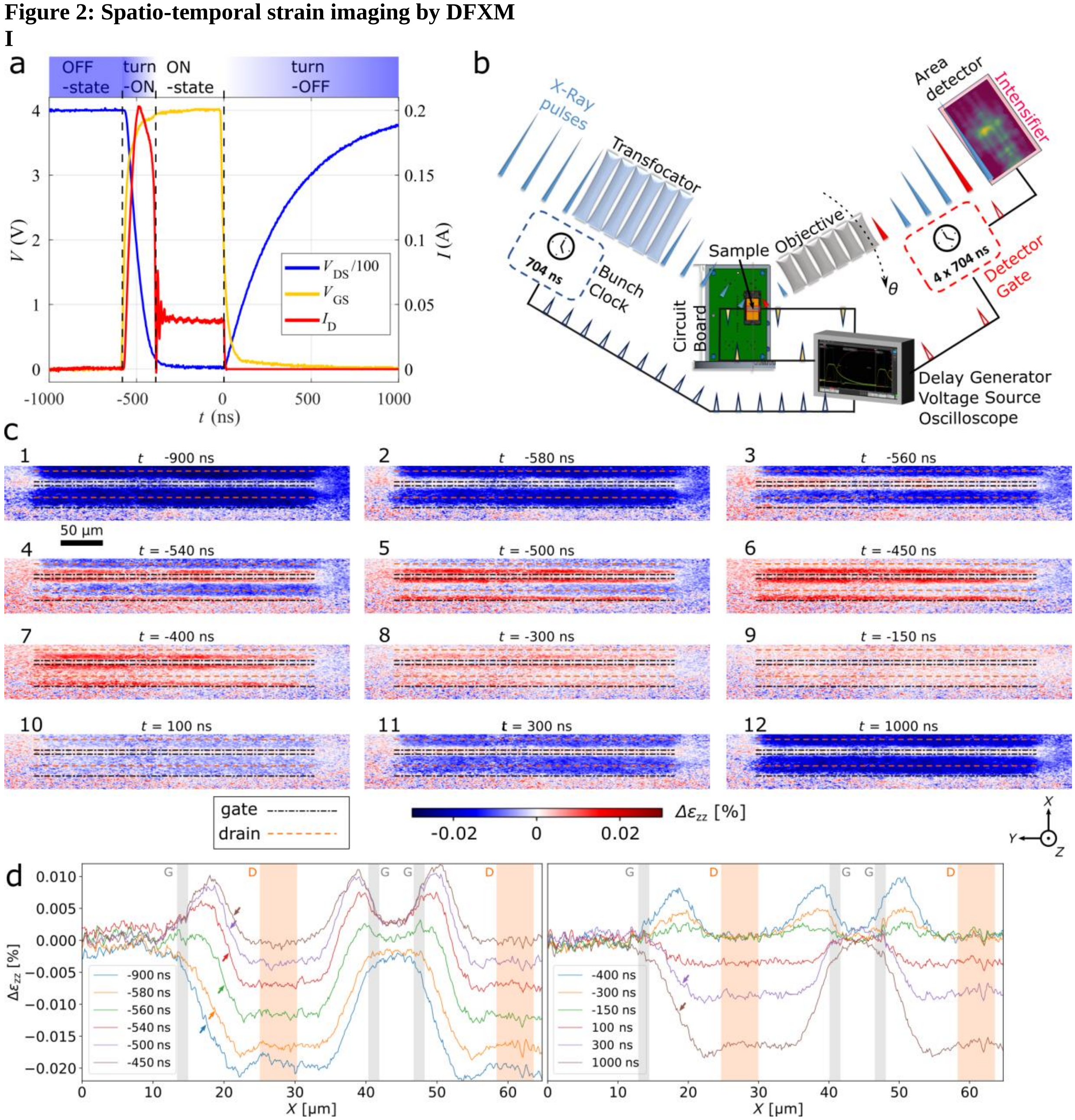

**a,** Time-dependent waveforms of drain-source voltage $V_{DS}(t)$, gate-source voltage $V_{GS}(t)$, and drain current $I_D(t)$ during switching. **b,** Setup for stroboscopic DFXM. **c,** Spatial images of the dynamic strain $\Delta\varepsilon_{zz}(t) = \varepsilon_{zz}(t) - \varepsilon_{zz}(t = 0)$ for 12 points in time, all panels share the same scale and color bars, the dashed lines indicate the center of gate and drain electrodes. **d,** Dynamic strain profiles $\Delta\varepsilon_{zz}(x)$ across the entire HEMT at the same points in time, taken from the strain images, the shading indicates the extent of gate (G) and drain (D) electrodes.

modulation $\Delta\varepsilon_{\mathrm{zz,IPE}}$ by the piezoelectric coefficients $d_{31}$, $d_{33}$ and Poisson number $\nu_{13}$, accounting for the Poisson effect [22]:

$$\Delta\varepsilon_{\mathrm{zz,IPE}} = \left(d_{33} - \frac{2d_{31}}{\frac{1}{\nu_{13}} - 1}\right)E_{\mathrm{z}} \tag{3}$$

Using literature values of $d_{31}$ = -1.4 pm/V, $d_{33}$ = 2.7 pm/V [23], and $\nu_{13}$ = 0.183 for GaN [24], compressive strain can be directly related to a downwards-pointing electric field of $E_{\mathrm{z}} \approx -0.6\ \frac{\mathrm{MV}}{\mathrm{cm}}$ in the gate-drain region during OFF-state ($t$ < -580 ns, **Fig. 2c1**). Through the course of the switching, the magnitude of the strain, and thus the corresponding electric field, follows the trend of $V_{\mathrm{DS}}(t)$. When turning ON at $t$ > -580 ns (**c2**), the compressive strain field reduces (**c3-6**), disappearing in the ON-state (**c7-9**), and rebuilding gradually when switching OFF again after $t$ > 0 ns (**c10-12**).

In the turn-ON phase, tensile strain is generated along the drain-sided gate edge (**Fig. 2c3-4**), reaching a maximum value of $\Delta\varepsilon_{\mathrm{zz,T}} \approx +1.5*10^{-4}$ between $t$ = -500 ns and $t$ = -450 ns (**c5-6**). This tension cannot be caused by the IPE, since it would require switching to an upwards-pointing electric field, which is an unreasonable assumption in the device. Instead, this strain field is attributed to thermal expansion due to self-heating during turn-ON [25]. During this period, $I_{\mathrm{D}}$ strongly increases, while simultaneously the voltage has not yet settled into its small ON-state value, resulting in a strong voltage-current overlap and an associated dissipation of electrical power $P(t)$ [26, 27]:

$$P = \int_{t_{\mathrm{OFF}}}^{t_{\mathrm{ON}}} V_{\mathrm{DS}}(t)\mathrm{I}_{\mathrm{D}}(t)dt \tag{4}$$

In the ON-state, electrical power dissipation is small due to the low $V_{\mathrm{DS}}$ and the hot areas are passively cooled via thermal conduction (**c7-9**), with the thermal strain disappearing after $t > 100$ ns (**c10-12**), as the active region approaches thermal equilibrium.

The translation of thermal strain to temperature $T$ is complex for spatially inhomogeneous distributions and may not be easily expressed by an analytic equation [28]. Moreover, the study of spatial strain profiles $\Delta\varepsilon_{\mathrm{zz}}(x)$ across the HEMT, plotted in **Fig. 2d**, reveals gradients from each gate towards the corresponding drain electrodes during OFF-state, turn-ON and turn-OFF, e.g. in the region indicated by the small arrows.

This indicates a gradient in the electric field and/or temperature at the gate edge, warranting a detailed consideration of localized potential and heating. A powerful means in the interpretation of *operando* strain data is the comparison of mechanical models to the experiment [29, 30], thus TCAD simulations of the device are performed, using as input the material parameters and measured time-dependent *I-V* characteristics.

We note that at all times, the strain evolution appears identical within each of the four parallel regions of the HEMT. Furthermore, there are no significant gradients of the differential strain along the electrodes, indicating uniform electric potential and temperature in $y$-direction $\left(\frac{\partial\Delta\varepsilon_{\mathrm{zz}}}{\partial \mathrm{y}} \approx 0\right)$. Thus, all relevant electromechanical processes may be reproduced in a model in cross-section view ($xz$-plane) of a single gate-drain region, with $x$ perpendicular to the electrodes and $z$ representing the depth.**Fig. 3a** shows simulated spatial distributions of the electric potential at four points in time during switching: OFF-state ($t$ < -580 ns); maximum power dissipation in turn-ON ($t$ = -490 ns); end of ON-state ($t$ = 0); turn-OFF ($t$ = 1000 ns). In the OFF-state, in the area between gate and drain there are lateral and depth-dependent potential gradients, reaching the maximum of $V_{DS}$ = 400 V underneath the drain contact. During turn-ON, the potential reduces in the active region, approaching a negligible value in the ON-state and recovering the original spatial distribution during the turn-OFF.

**Figure 3: Modelling of electrical and thermal dynamics**

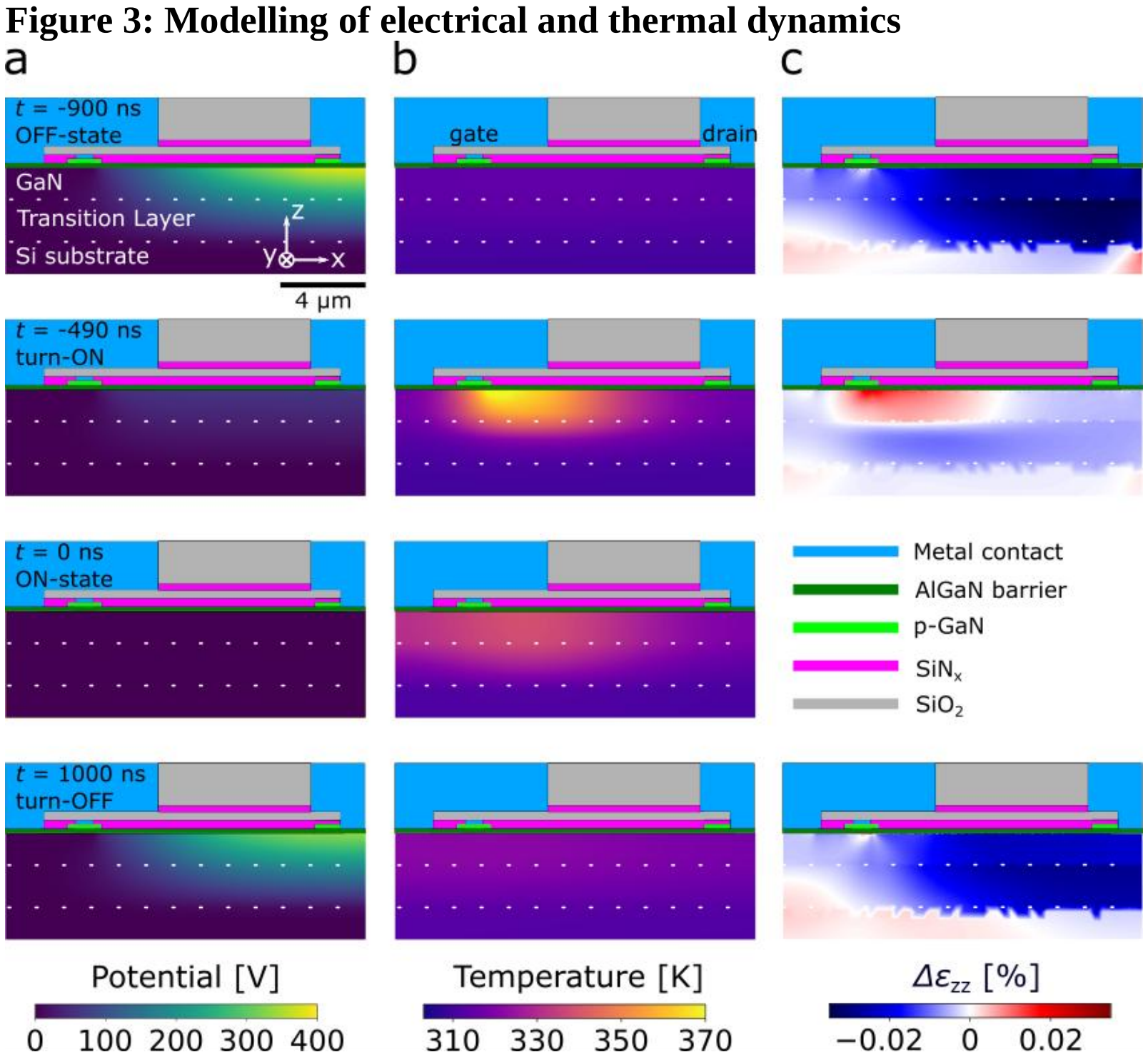

**a,** TCAD simulations of electric potential $V$ in cross-section of the HEMT for four points in time. **b,** Temperature $T$ in the same model. **c,** FEM simulation of the differential lattice strain $\Delta\varepsilon_{zz}(t)$.

Furthermore, the TCAD model incorporates electrothermal dynamics, providing temperature distributions in the device, presented in **Fig. 3b**. In the OFF-state, the temperature is homogenous. During turn-ON, a thermal hotspot occurs at the surface, near the drain-side edge of the gate electrode, with a maximum heating of $\Delta T \approx +70$ K. The temperature in the (Al)GaN layers decreases with distance to the hotspot, which also dissipates over time, having reduced to $\Delta T < 30$ K at the end of the ON-state, and disappearing during turn-OFF.

To obtain the same physical quantity as probed by the experiment, a FEM model in *COMSOL Multiphysics* was employed to translate electric field and temperature distributions from TCAD into the differential strain $\Delta\varepsilon_{zz}\,(t, x, y)$, with the definition from Eq. (1). Images of the individual contributions from IPE $\Delta\varepsilon_{zz,IPE}$ and thermal strain $\Delta\varepsilon_{zz,T}$ are provided in Supporting Figure 6, their sum is presented in **Fig. 3c**. The combined TCAD and FEM simulations predict that GaN and transition layer are under compressive strain between gate and drain during OFF-state and turn-OFF due to the downwards potential gradient.

However, in the turn-ON, the thermal hotspot generates tensile strain in the GaN layer by the gate. Interestingly, the transition layer is predicted to become more compressive underneath the hotspot instead, due to the pressure exerted downwards as the upper layers expand.

**Figure 4: Spatio-temporal strain distributions**

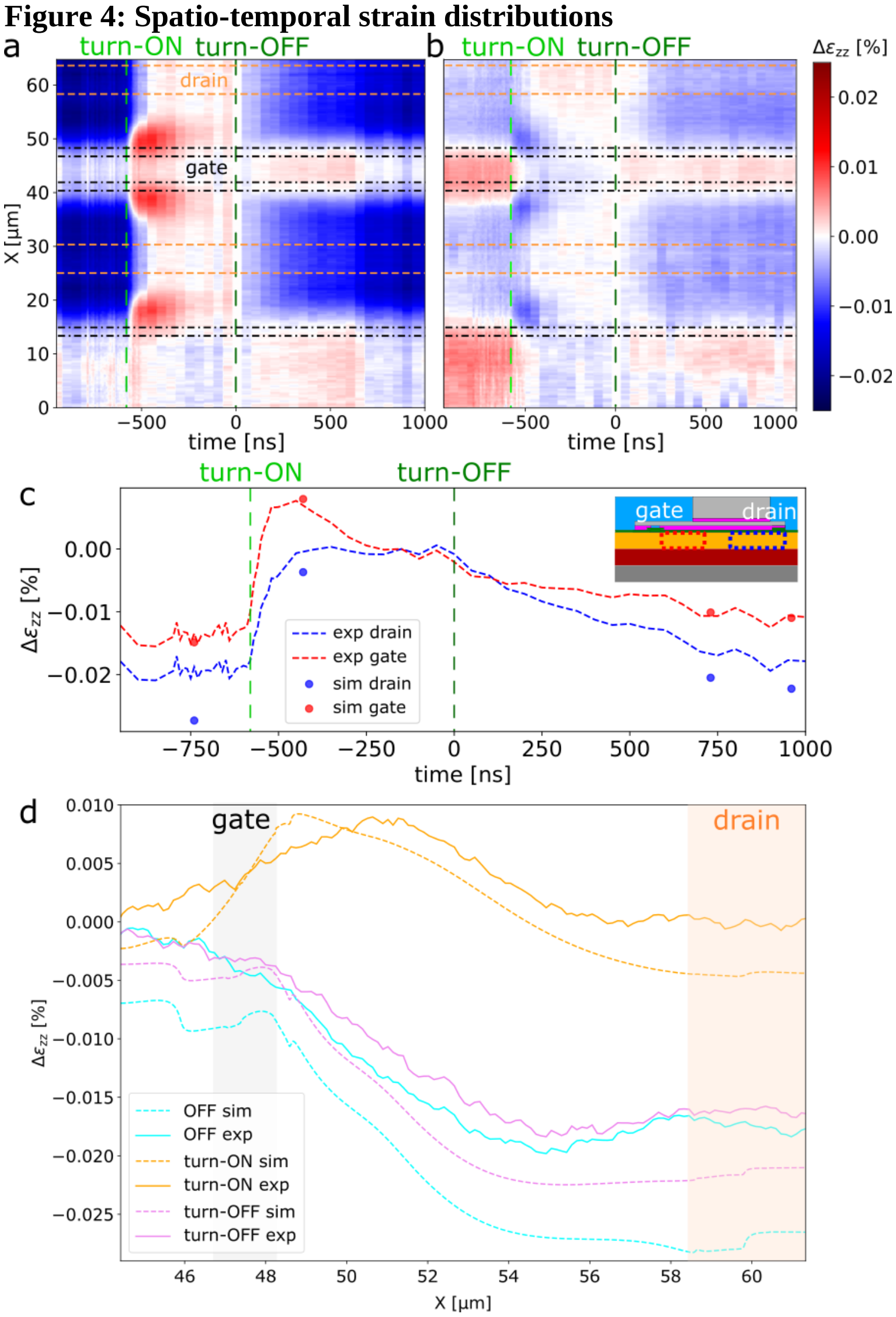

**a**, 2D distribution of dynamic lattice strain $\varDelta\varepsilon_{\mathrm{zz}}(t,x)$ in the GaN layers. **b**, The same plot for the transition layer. **c**, Comparison of temporal profiles in experiment and simulation for the GaN layer, the inset indicates the spatial sections used for averaging. **d**, Spatial curves of experimental and simulated differential strain for OFF-state, turn-ON and turn-OFF, the extent of the electrodes is indicated by the shading.

For quantitative comparison of experiment and modelling, the 3D DFXM dataset $\varDelta\varepsilon_{\mathrm{zz}}(t,x,y)$ is reduced by averaging along *y*-direction, yielding the 2D spatio-temporal distribution of differential strain $\varDelta\varepsilon_{\mathrm{zz}}(t,x)$ plotted in **Fig. 4a**. Remarkably, the separation of IPE and heating is generally straightforward from this visualization due to the fine spatio-temporal resolution of the experimental data, as it is apparent that the two strain fields $\varDelta\varepsilon_{\mathrm{zz,IPE}}(t,x,y) < 0$ and $\varDelta\varepsilon_{\mathrm{zz,T}}(t,x,y) > 0$ only overlap close to the edge of the gate electrodes around $t \approx$ -500 ns. In this specific region of spacetime, $\varDelta\varepsilon_{\mathrm{zz}}$ is tensile, indicating that the thermal contribution is dominant. Moreover, we observe that the tensile section widens during the first ~50 ns of turn-ON until reaching a maximum width of ~5 µm, presumably through thermal diffusion from the hotspot. A plot of $\varDelta\varepsilon_{\mathrm{zz}}(t,x)$ in the transition layer of the heterostructure (see Fig.1b) is presented in **Fig. 4b** (strain images are provided in Supporting Figure 7). Indeed, we find in the experiment that close to the gate edge, the dynamic strain during turn-ON is compressive rather than tensile, as predicted by the modelling.

To capture the evolution of IPE and heating, in **Fig. 4c** two temporal profiles of the differential strain $\varDelta\varepsilon_{\mathrm{zz}}(t)$ are extracted from the DFXM data of the GaN layer in 4 and 5 µm-wide slices adjacent to gate and drain, respectively, as indicated by the inset. The FEM simulations, represented by temporal datapoints obtained from the same spatial sections in the model, are in excellent agreement with the experiment for the strain evolution in the gate section, where the thermal strain dominates. In the drain region, which is affected almost exclusively by the IPE, the experimental and simulated strain profiles follow the same temporal trend, but the total magnitude is weaker by ~20% for the former. This difference is well within the spread of the relevant coupling parameters $d_{31}$, $d_{33}$ and $\nu_{13}$ (see Eq. 2) documented in the literature [24, 23].
To study the spatial shape of the dynamic strain fields between gate and drain, experimental and simulated profiles of $\Delta\varepsilon_{zz}(x)$ are compared in **Fig. 4d**. During OFF state (cyan curve), the 2DEG is depleted and the region between gate, drain and substrate is highly resistive. Here, the electrostatic potential decreases laterally between drain and gate and vertically between drain and substrate. Since the piezoelectric strain is proportional to the vertical electric field, this generates a lateral gradient from the gate edge towards a plateau by the drain. The corresponding linear decrease of $\Delta\varepsilon_{zz}(x)$ by ~0.015 % over ~6 µm is reproduced in both experiment and simulation.
However, there is a rigid baseline offset between the two curves, which we attribute to gradual global heat accumulation in the device over many cycles. In the experiment, each data point is obtained by averaging over >$10^5$ pulses, while the simulation captures the sample after only six full cycles, i.e. at the onset of its operation. We attribute the baseline shift of the dynamic strain to per-pulse global heating of the model.
During turn-ON (orange curve), we find a local maximum in $\Delta\varepsilon_{zz}(x)$, which is well matched in amplitude between experiment and simulation, indicating that the model prediction for a maximum heating of $\Delta T \approx +70$ K is accurate. However, in the experiment the spatial profile of thermal strain is symmetric with its maximum at ~2 µm distance from the gate edge at the rim of the metal pad, while the simulation predicts the strongest deformation directly underneath the gate edge with a shoulder towards the drain side. This difference might be caused by a gradual charge redistribution in the carbon-doped layer during many OFF/ON cycles, which can modify the electrostatic potential near the gate [31], shifting the thermal hotspot towards the drain. This phenomenon may not be significant during the small number of switching cycles incorporated in the TCAD simulation.

# Discussion

Our operando X-ray diffraction microscopy measurements provide direct access to dynamic electromechanical and thermal strain fields inside a high-power transistor during 355 kHz operation. This mechanical insight translates into the spatio-temporal evolution of internal fields, revealing field gradients approaching −0.6 MV cm$^{-1}$ and a transient ~70 K hotspot a few microns wide at the gate edge. These phenomena may be conceptually predicted by modeling, but have not been validated at the relevant spatio-temporal scales. This experimental insight proved crucial, as the hotspot is shifted relative to theoretical expectations.
Visualizing these dynamics in situ provides new insight into the mechanisms limiting performance and reliability of power electronics. The field dynamics inferred from the strain transients modulate the two-dimensional electron gas, promote hot-carrier injection and trapping, and contribute to drift in the ON-state resistance $R_{DS,on}$ [32]. The thermal hotspot may reduce mobility through increased phonon scattering [33], and amplify conduction and switching losses. Together, these insights highlight the importance of field-shaping and thermal-management strategies, e.g., in the design of gate and field-plate geometries [34, 35], to suppress charge trapping and enhance material robustness.
More broadly, our results demonstrate that operando X-ray microscopy establishes a direct connection between device-scale electrical measurements and the underlying nanoscale structural processes. The brilliance and ultrashort pulses of next-generation synchrotrons and X-ray free-electron lasers will push time-resolved diffraction microscopy into the picosecond domain [36], enabling comprehensive imaging of piezoelectric and thermal strain landscapes in a wide range of functional materials. This capability will open new avenues for understanding transient mechanical phenomena central to technologies such as memristors, ultrasonic actuators, and ultrafast photodiodes, establishing operando structural imaging as a foundational tool for next-generation microelectronics and beyond.

# Experimental section

## Electric stress testing

The GaN HEMT test circuit for hard-switching with a capacitive load is represented in **Supporting Fig. 1**. The value of the load resistor defines the ON-state current and therefore the ON-state conduction heating, which was chosen as $R_{\mathrm{load}}$ =10 kΩ. The load capacitor is given by the inherent parasitic capacitance of the Printed Circuit Board (PCB) and socket, with a value of $C_{\mathrm{load}}$ = 34 pF. During the synchrotron experiment, the $V_{\mathrm{GS}}(t)$ and $V_{\mathrm{DS}}(t)$ curves of the GaN HEMT were measured with differential high-voltage probes and recorded together with the trigger pulses from the delay generator by a digital oscilloscope to ensure accurate pump-probe synchronization. The drain current $I_{\mathrm{D}}$ was measured separately and matched to the experimental data afterwards based on the voltage characteristics.

## Dark Field X-ray Diffraction Microscopy

The stroboscopic DFXM experiment was carried out at beamline ID01 of the European Synchrotron Radiation Facility (ESRF), with the setup sketched in **Fig. 2b**. The X-ray energy was set to 8.92 keV and sample and detector were aligned for the 0002 Bragg reflection of GaN in a horizontal scattering geometry. The objective comprised 50 Be CRLs with a curvature radius of 50 µm with a focal length $f \approx 12$ mm. The data was recorded with an *Andor Zyla 5.5* area detector with 2160 x 2560 pixels with a pixel size of 6.5 µm placed inside an evacuated flight tube at a distance of $d \approx 6.5$ m to the objective, yielding in a magnification factor of $\frac{d}{f} \approx 54$. Due to the oblique diffraction angle of $\theta \approx 15.55$ °, the effective pixel size on the detector broadens by a factor of $\sin(\theta)^{-1} \approx 3.73$ within the diffraction plane, resulting in a spatial resolution of 119 nm along $x$-direction and 444 nm along $y$. To optimize the resolution across the active channel, the electrodes were oriented parallel to the diffraction plane.
In DFXM geometry, a real space image of the GaN layers in diffraction contrast could be obtained with an exposure time of only a few seconds and a Field of View (FOV) of several hundred µm$^2$. The combination of large FOV and fast acquisition time is crucial when exploring a large spatio-temporal parameter space to localize a dynamic effect.
The lattice strain is scanned by rotating the objective to obtain a curve of $I(\theta)$ in every point on the sample, from which the local out-of-plane lattice parameter $c$ and the corresponding strain tensor component $\varepsilon_{\mathrm{zz}}$ are calculated by re-arranging the Bragg equation for the 0002 reflection:

$$c = \frac{\lambda}{\sin(\theta)} \quad (5)$$

$$\varepsilon_{zz} = \frac{c}{c_{\mathrm{GaN}}} - 1 \quad (6)$$

With the X-ray wavelength $\lambda = 0.139$ nm and the literature value for the lattice parameter of GaN $c_{\mathrm{GaN}} = 0.5186$ nm [47].
In the 4bunch-mode of the ESRF, the synchrotron beam consists of pulses with a frequency of 1.42 MHz and duration of ~40 ps. The electrical switching was synchronized to the X-ray pulses with a *Stanford Instruments DGS 645*, by which the delay time between the electrical signal for switching the HEMT and the X-ray probe pulse was scanned for dynamic strain imaging. Since switching was performed at exactly one quarter of the X-ray frequency, the detector was gated with an intensifier to select every fourth pulse, ensuring that each data point comprises only X-rays scattered from the sample at a specific delay time *t*.

## TCAD modeling

TCAD simulation is performed with the *Synopsys Sentaurus* software, employing a thermodynamic model for electron and hole current densities to obtain a self-consistent solution of HEMT operation including self-heating. The model is calibrated to temperature-dependent DC and AC characteristics with thermal conductivities taken from Refs. [45, 48]. Carbon doping is implemented as a dominant acceptor model [49]. The GaN HEMT test circuit is simulated and the load capacitance is adapted to account for parasitic capacitances in the experiment and match the drain current transient during turn-ON. Initially, the temperature in the model is set to be spatially homogenous at 300 K.
Six full OFF/ON switching cycles are simulated, during which the device experiences fast transient self-heating effects and a slow cumulative self-heating to ~315K over all six pulses, after which the simulation is stopped due to processing time constraints. The electrostatic potential and lattice temperature distributions during the last cycle are exported for FEM modelling.

## COMSOL modeling

The IPE, i.e. deformation of a material under an applied potential difference, is simulated in a finite element framework with *COMSOL Multiphysics*. The geometry of the TCAD simulation is re-meshed with quad elements with quadratic shape functions. Symmetry boundary conditions are applied on both sides of the 2D model, the lowest, leftmost node is constrained in both directions, while the lowest rightmost is allowed to move in one direction, allowing for an unconstrained expansion of the system.

The TCAD inputs are imported as an interpolation function of potential and temperature in space and time. For mechanical properties of the wurtzite (Al)GaN structure, elastic constants and piezoelectric coupling matrix from Ref. [50] and coefficients of thermal expansion from Ref. [51] were utilized. A small strain setting is employed, where the total strain comprises thermal, piezoelectric and elastic components.

**Acknowledgments:** We acknowledge the European Synchrotron Radiation Facility (ESRF) for provision of synchrotron radiation facilities under proposal number MA-6231 and thank the staff for assistance in using beamline ID01. Furthermore, we thank Dr. Guillaume Gaisne for his support on FAIR data treatment and maintaining the NOMAD repository. C. Corley-Wiciak and M. Petersmann thank Dr. Costanza Manganelli for her advice on COMSOL modelling.

**Funding:** This study has received financial support from the European Union: AddMorePower (GA 101091621).

**Competing interests:**

The authors have no competing interests to disclose.

**Data, code, and materials availability:**

The datasets generated and analyzed during the current study are available in the ESRF data archive (https://doi.org/10.15151/ESRF-ES-1854292272). Derived data are available from the AddMorePower Nomad Repository (10.17172/nomad.natt-46vn). All materials used in this study are described in the materials and methods.

# Supplementary Information

Details on the Electrical Setup

The circuits for two electrical setups to measure different types of the drain current are illustrated in **S. Fig. 1**: (**a**) The drain current $I_{Load}$ from the beamline setup, which includes the load current that does not pass through the transistor but instead charges the capacitor $C_{load}$; (**b**) the actual drain current $I_D$ through the transistor, obtained from a separate measurement under a comparable wafer-level switching condition.

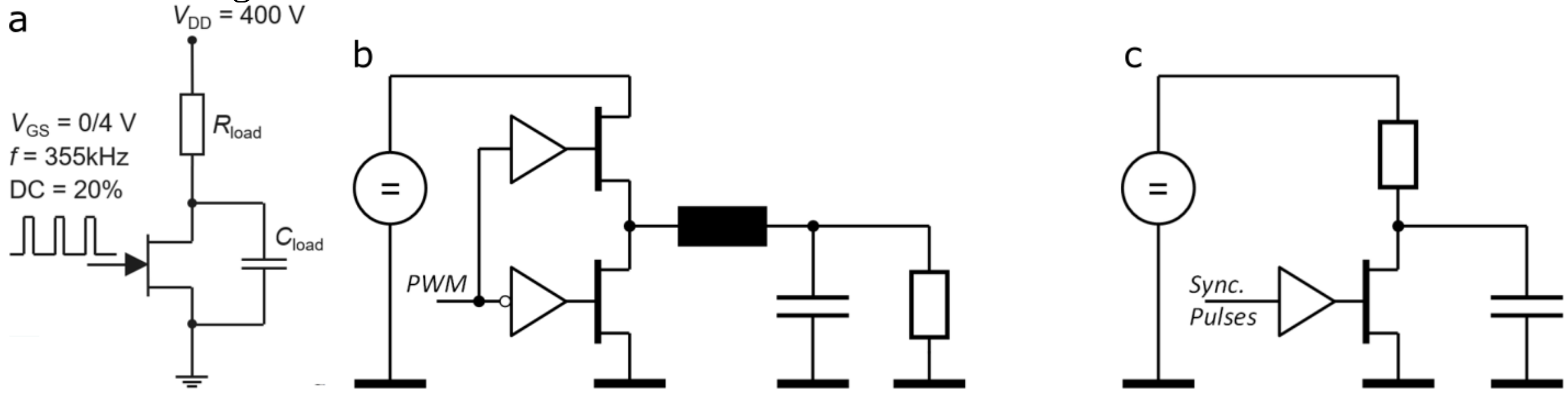

**Supplementary Fig. 1: a**, Schematics of the HEMT test circuit for hard-switching with capacitive load; **b,** HEMT application circuit; **c**, simplified test circuit for hard turn-ON.

Performing *operando* X-ray microscopy on the GaN HEMT imposes certain requirements on the electrical test setup: The switching instants must be accurately synchronized with the X-Ray pulse sequence, which is determined by the bunch frequency of the synchrotron. Moreover, static self-heating of the sample by excessive power dissipation should be minimized. Therefore, instead of true Pulse-Width Modulation (PWM) operation with an inductive load current, as in a DC converter application, a simplified stress test circuit was chosen (**panel c**), that models solely the hard turn-on of the GaN HEMT channel against a capacitor that represents the parasitic circuit load impedance of the switching node. The capacitor is recharged between pulses by a high ohmic resistor, thus the static drain current of the tested device remains sufficiently low to avoid self-heating, while the switching stress may be adjusted by changing the capacitor value as well as the gate drive voltage amplitude.

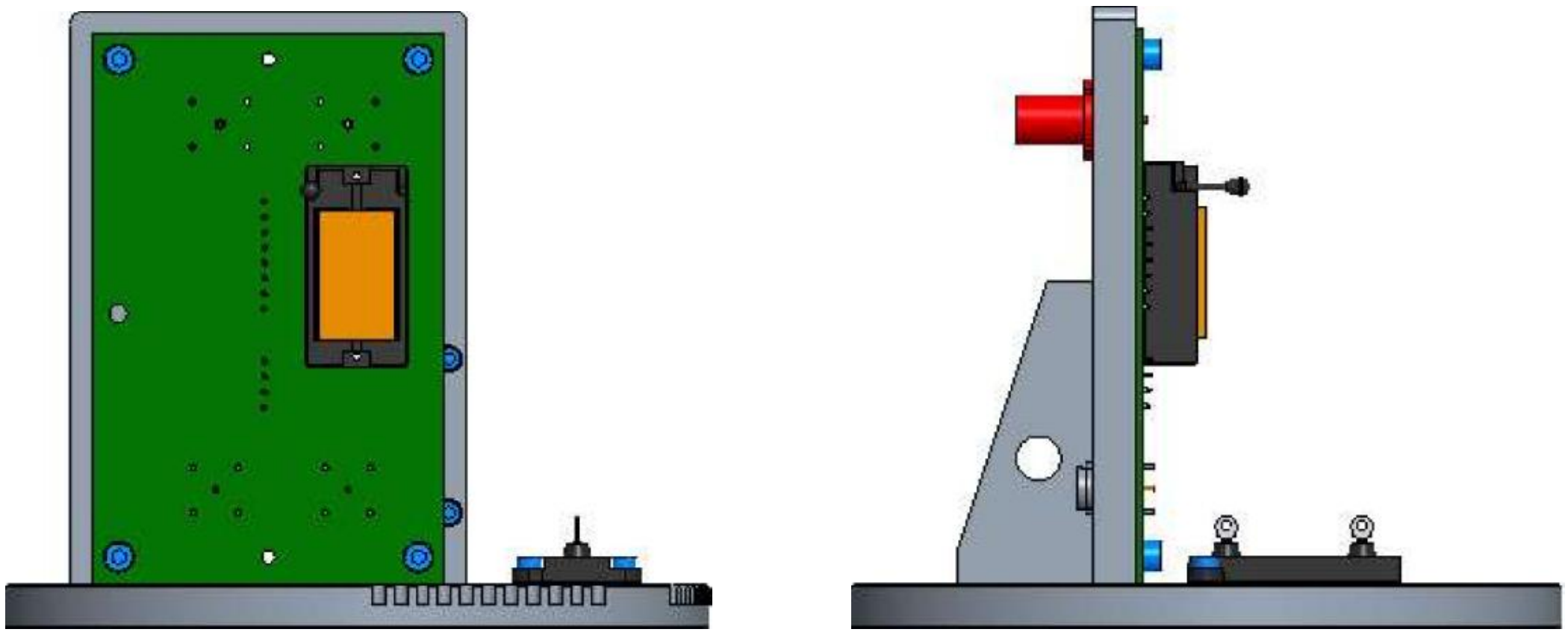

**Supplementary Figure 2:** Mechanical fixture and carrier PCB with ZIF socket to support the GaN HEMT sample during the synchrotron experiment.

The HEMT specimen was bonded in an open ceramic package to provide free access to the chip. The package was securely mounted in a standard Zero Insertion Force (ZIF) socket soldered to a PCB together with a fast gate driver and the required passive components. A purpose-built mechanical fixture, sketched in **S. Fig. 2**, was mounted on an adjustable pod to support the PCB and socket, holding the chip securely but mobile on the beamline diffractometer. The re-charge resistor was placed on the metal frame for optimum cooling to minimize heating of the sample.

Strain Data in the GaN layers

Images of the strain $\varepsilon_{zz}$ in the GaN layers are presented in **S. Fig. 3**. In each frame, there are pronounced line features along the edges of the metallic contacts (gate/source and drain), indicating they are subject to some amount of internal stress [24]. The local strain modulation in the GaN layers originating from the drain electrodes is compressive with $\varepsilon_{zz} \approx -10^{-4}$, while that of the gate is tensile ($\varepsilon_{zz} \approx +1.5 * 10^{-4}$). This may be attributed to differences in either the deposition conditions of the material, or the contact geometry, which determines the static strain

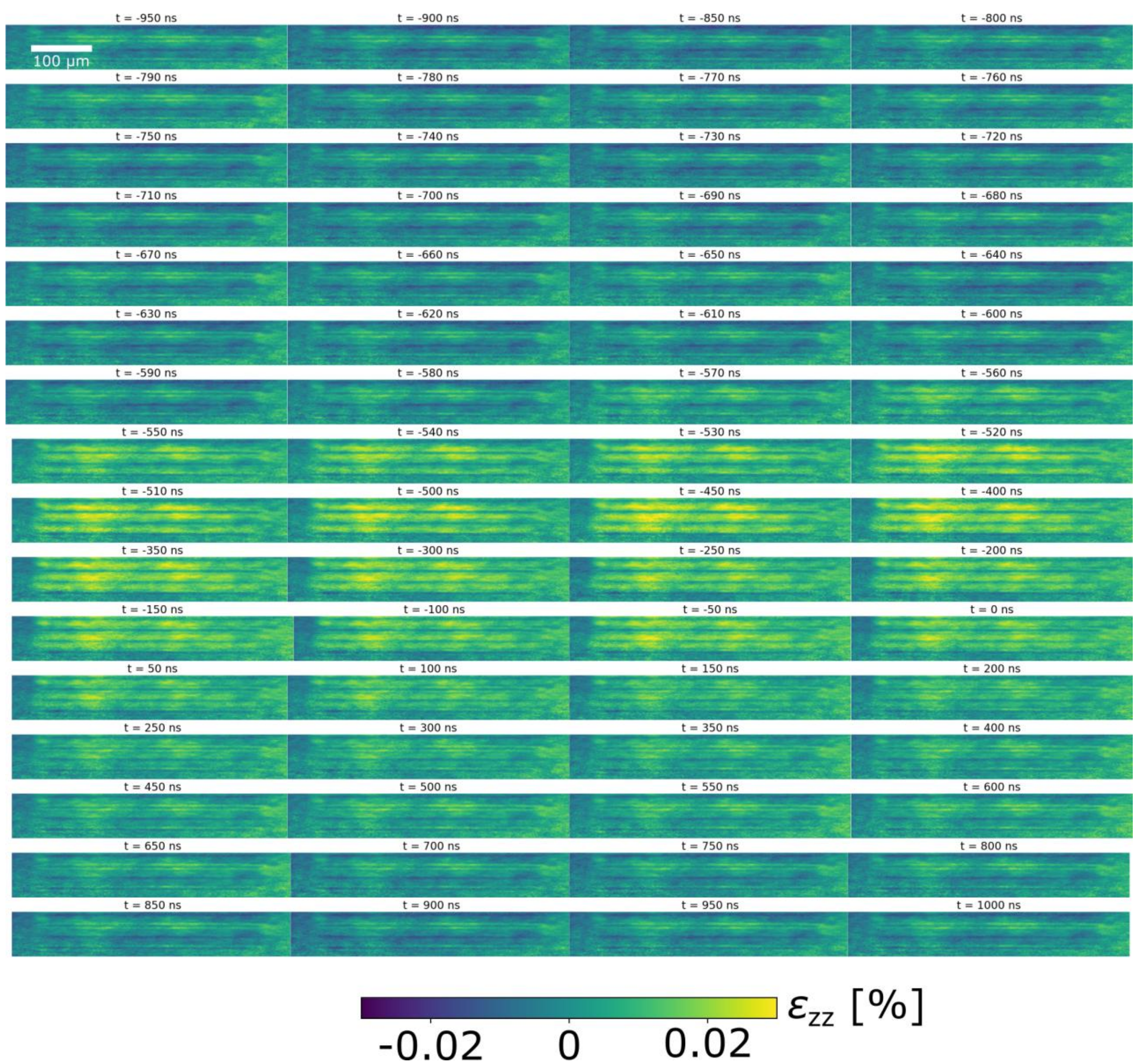

**Supplementary Figure 3:** Spatial images of the instantaneous strain $\varepsilon_{zz}$ in the GaN layers for 65 points in time, all images share the same scale and color bars.

field induced by the stressed metal stripes in the epilayers [51].

From Eq. (1) in the main text, the spatial distribution of the dynamic strain $\Delta\varepsilon_{zz}$ is calculated at each point in time, yielding the series of 64 images in **S. Fig. 4,** excluding $\Delta\varepsilon_{zz}(t = 0) = 0$.

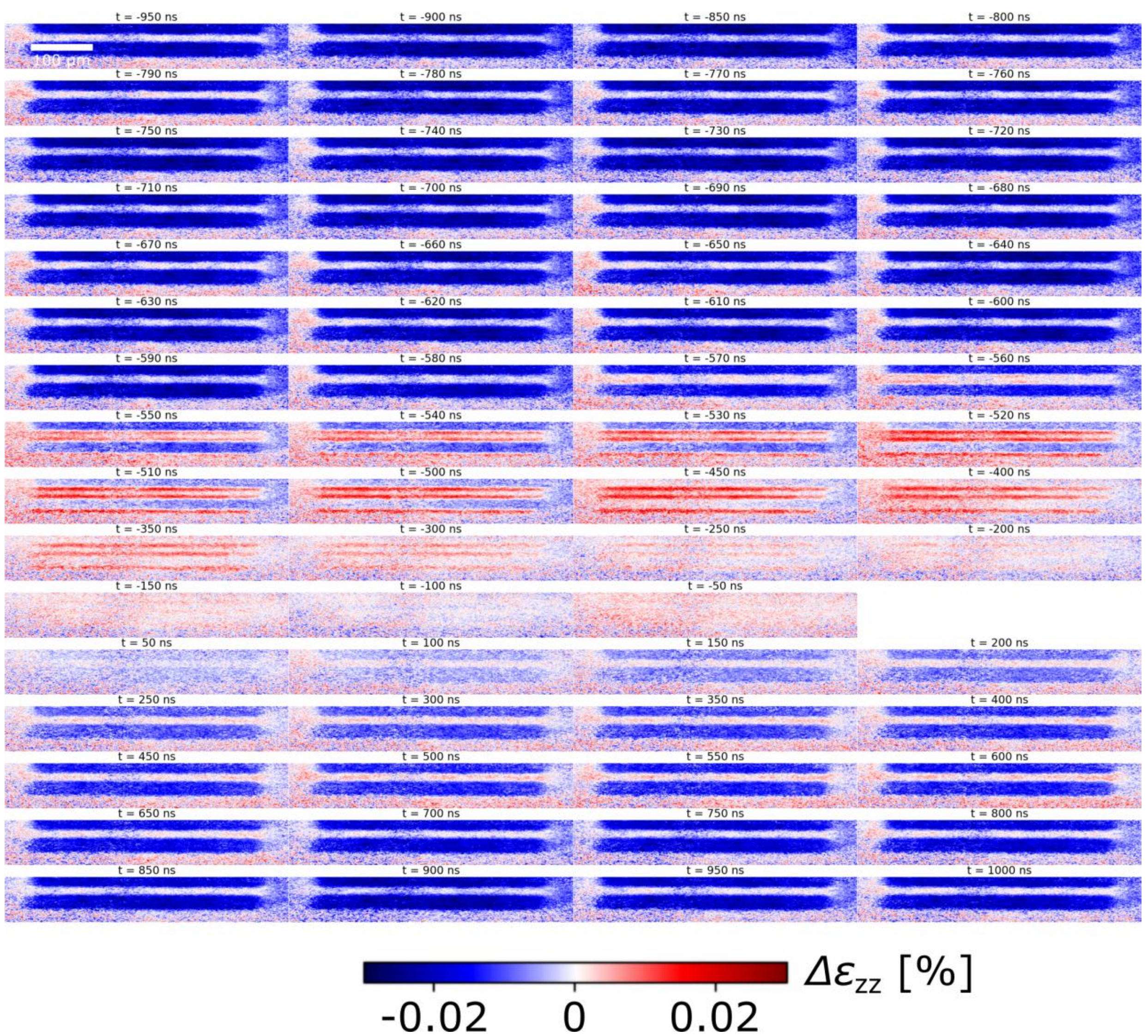

**Supplementary Fig. 4:** Spatial images of $\Delta\varepsilon_{zz}(t)$ in the GaN layers for 64 points in time, all images share the same scale and color bars.

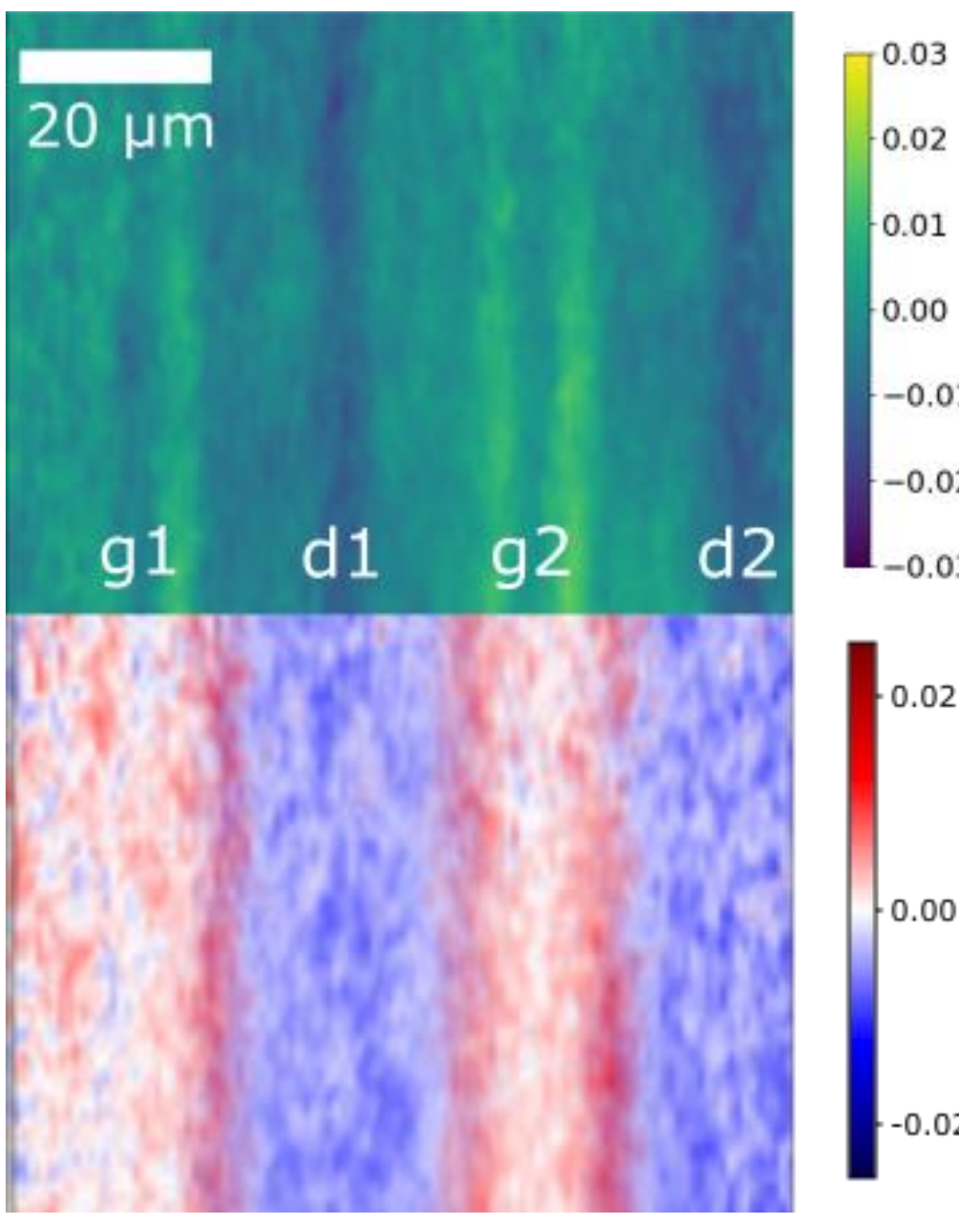

**Supplementary Figure 5:** Side-by-side comparison of static strain $\varepsilon_{zz}$ and dynamic strain $\Delta\varepsilon_{zz}$ in the GaN layer

In a side-by-side comparison of sections of $\varepsilon_{zz}$ and $\Delta\varepsilon_{zz}$ in the HEMT, shown by **S. Fig. 5**, it is evident that they exhibit different spatial features. The static strain is dominated by the internal stress at the edges of the TiN electrodes, which appears time-independent, as it is not present in the dynamic strain image. Thus, we conclude that no additional stress is generated in the metallic electrode material during switching.

Thermal and Electromechanical FEM Simulations

**S. Fig. 6** shows the dynamic strain in cross-section view of the HEMT due to the IPE $\Delta\varepsilon_{\mathrm{zz,IPE}}$ and local heating $\Delta\varepsilon_{\mathrm{zz,T}}$ (see Eq. (2) in the main text) calculated by the FEM model at four points in time, based on the spatial distribution of temperature and electric potential predicted by the TCAD model.

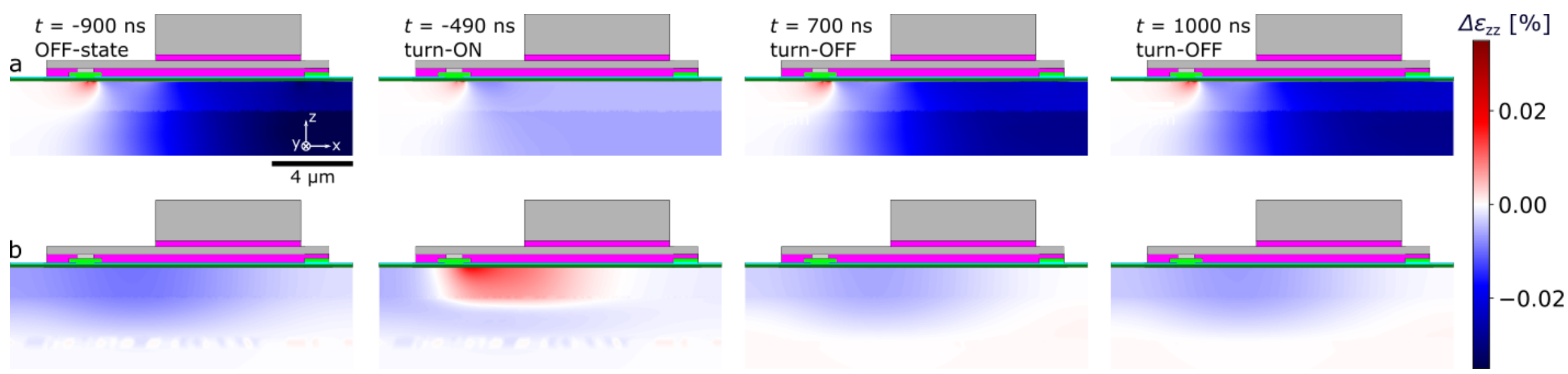

**Supplementary Figure 6:** FEM calculation of dynamic lattice strain due to (**a**) IPE $\Delta\varepsilon_{\mathrm{zz,IPE}}$ and (**b**) heating $\Delta\varepsilon_{\mathrm{zz,T}}$.

Strain Data in the transition layer

The diffraction curve measured by DFXM contained a peak for the AlGaN/AlN transition layer of the heterostructure, allowing to determine its dynamic strain by tracking its relative shift in the same way as for GaN layers. The resulting images of $\Delta\varepsilon_{zz}(t)$ in the transition layer are presented in **S. Fig. 7.**

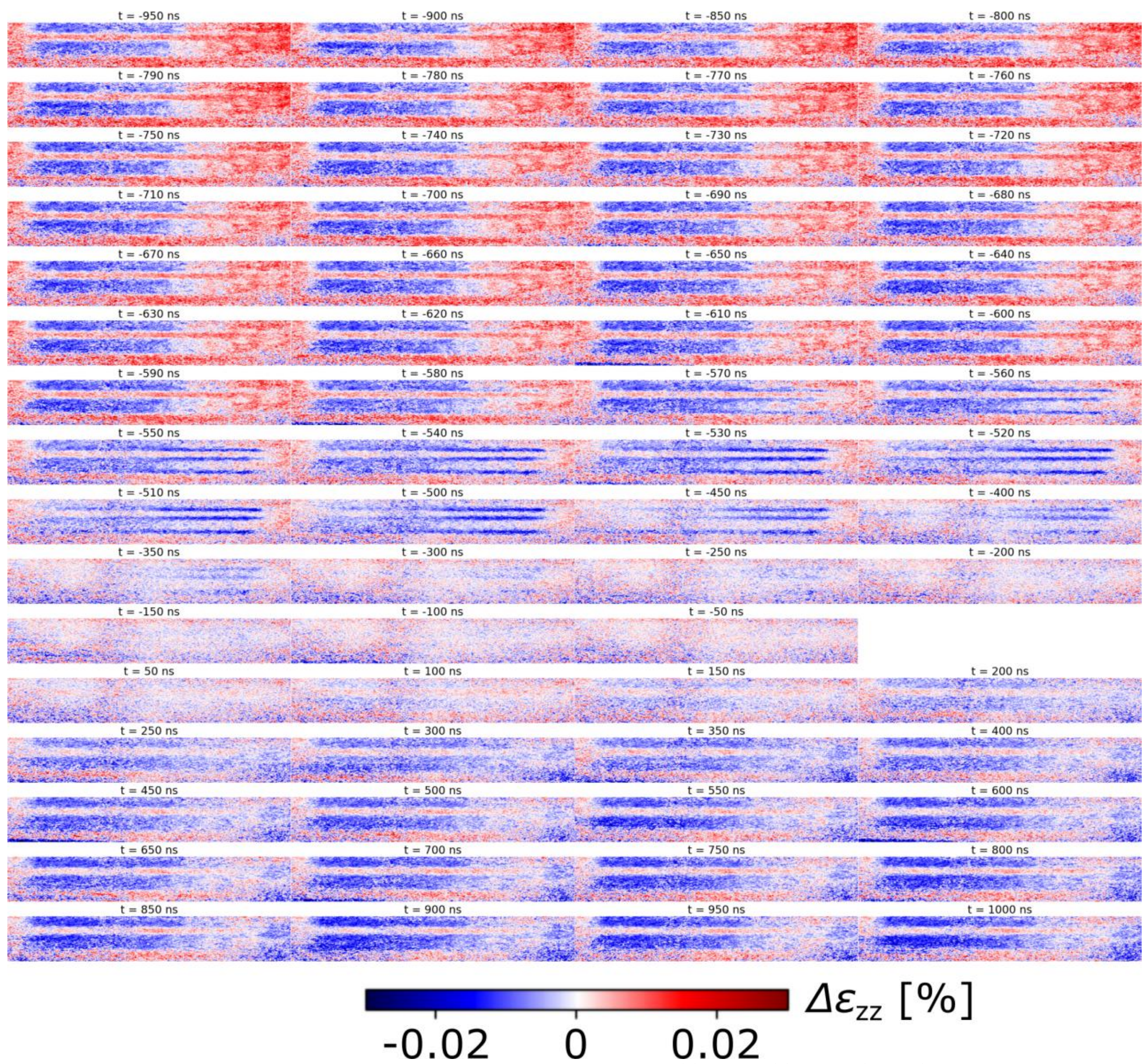

**Supplementary Figure 7:** Spatial images of the dynamic strain $\Delta\varepsilon_{zz}(t) = \varepsilon_{zz}(t) - \varepsilon_{zz}(t = 0)$ in the transition layer for 64 points in time, all images share the same scale and color bars.